\documentclass[aps,prl,reprint,superscriptaddress,showpacs]{revtex4-1}

\usepackage{color}

\usepackage{graphicx}
\usepackage{booktabs}
\usepackage{bm}
\usepackage{longtable}
\usepackage{amsmath}
\usepackage{dcolumn}
\usepackage{placeins}
\usepackage{expdlist}

\usepackage{siunitx}

\begin{document}

\title {Generation of atomic spin orientation with a linearly polarised beam in room-temperature alkali-metal vapour}

\author{P. Bevington} 
\affiliation{National Physical Laboratory, Hampton Road, Teddington TW11 0LW, United Kingdom}
\affiliation{Department of Physics, University of Strathclyde, Glasgow G4 0NG, United Kingdom}
\author{R.\, Gartman}
\affiliation{National Physical Laboratory, Hampton Road, Teddington TW11 0LW, United Kingdom}
\author{W.\, Chalupczak}
\affiliation{National Physical Laboratory, Hampton Road, Teddington TW11 0LW, United Kingdom}


\begin{abstract}
Traditionally, atomic spin orientation is achieved by the transfer of angular momentum from polarised light to an atomic system. We demonstrate the mechanism of orientation generation in room-temperature caesium vapours that combines three elements: optical pumping, non-linear spin dynamics and spin-exchange collisions. Through the variation of the spin-exchange relaxation rate, the transition between an aligned and an oriented atomic sample is presented. The observation is performed by monitoring the atomic radio-frequency spectra. 
The measurement configuration discussed, paves the way to simple and robust radio-frequency atomic magnetometers that are based on a single low power laser diode that approach the performance of  multi-laser pump-probe systems.

 \end{abstract}

\maketitle

\section{Introduction} 
The generation of spin polarization is an essential step in the study and application of a large variety of systems, from solid-state samples \cite{Awschalom2018} to cold atomic ensembles \cite{Jessen2013, Chalopin2018}. In the atomic physics domain, the standard method (optical pumping) relies on the transfer of angular momentum from polarised light to the atomic system \cite{Happer1972}. While a typical scheme involves the interaction of an atomic sample with a circularly polarised laser beam propagating along a static magnetic field, other configurations, including different polarisation \cite{Fortson1987, Klipstein1996, Andalkar2002} and number of lasers \cite{Wasilewski2010}, have been demonstrated. 
Optical pumping also covers the transfer of optical angular momentum to the target atoms achieved via spin-exchange collisions (SEC) \cite{Walker1997, Appelt1998}.  Another category of spin polarization processes combines optical pumping with non-linear spin dynamics \cite{Smith2004, Chalupczak2015}. One particular realization of this is the so-called effect of alignment to orientation conversion, which involves the evolution of an atomic population imbalance in mutually orthogonal magnetic and electric fields \cite{Cohen-Tannoudji1969, Lombardi1969, Hilborn1994, Budker2000, Kuntz2002, Auzinsh2010, Rochester2012}. In this way, tensor polarization, (alignment) where the spins are aligned along a preferred axis but no preferred direction, can be transformed into a vector polarization, (orientation) where spins are biased in one direction \cite{Auzinsh2010}. 
A similar efect has been achieved within the excited state hyperfine sublevels in the presence of either an electric or magnetic field \cite{Krainska1979, Han1991, Alnis2001, Auzinsh2006, Auzinsh2015}. 

\begin{figure}[h!]
\includegraphics[width=\columnwidth]{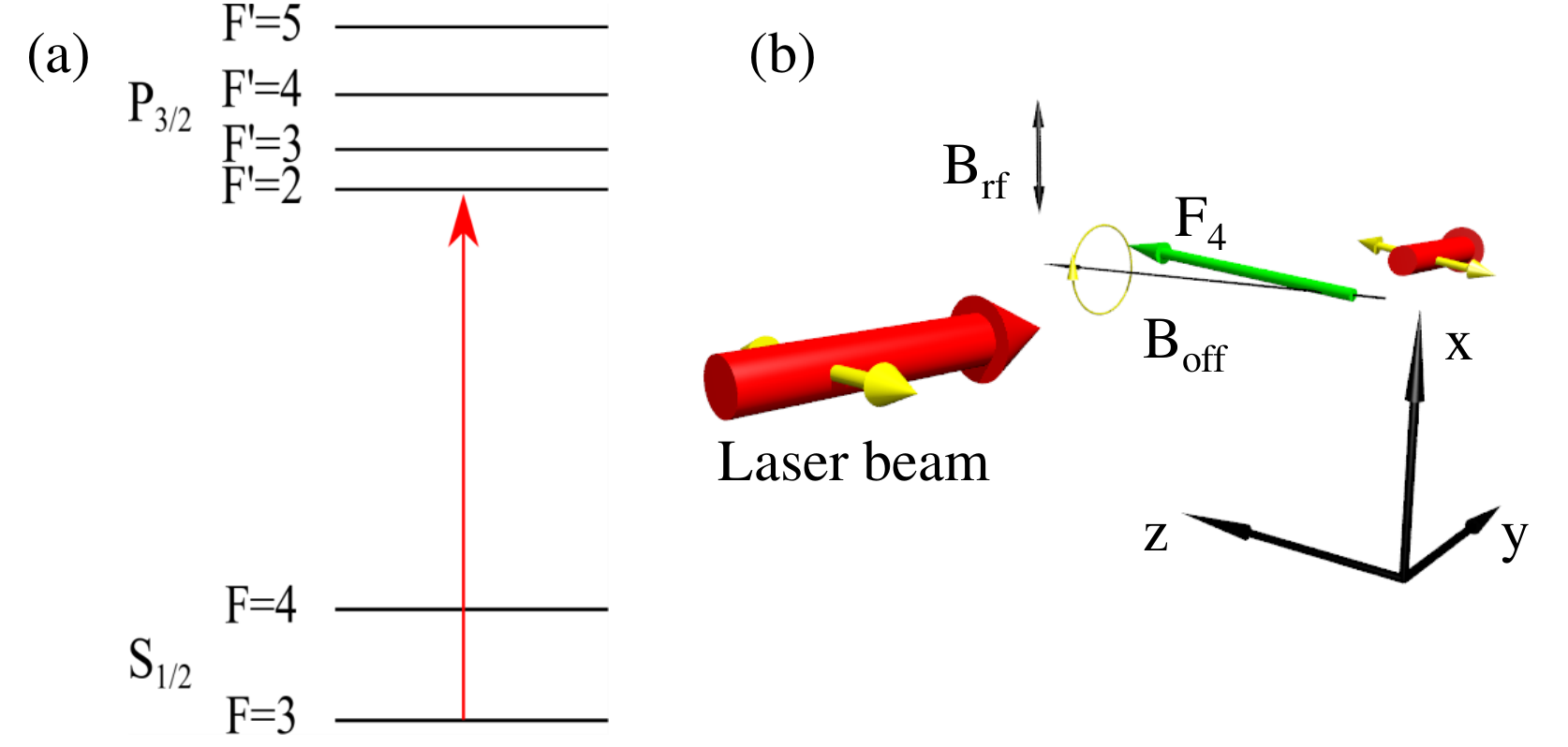}
\caption{(a) A linearly polarized laser beam, acting as the pump, transfers population between, and, creates a population imbalance within the F=3 and F=4 caesium ground state manifolds, i.e. atomic spin along the direction of $B_{off}$, marked with a black arrow in (b). (b) A weak radio-frequency field, $B_{rf}$, orthogonal to the $\hat{z}$ axis creates coherences between adjacent F=4 Zeeman sublevels, causing the atomic spin polarisation to precess around the $B_{off}$. The same linearly polarized laser beam, acting as the probe, monitors the spin precession via the Faraday effect. In some measurements, a weak circularly polarised beam propagating along the $\hat{z}$ axis is used to modify the population imbalance created by the linearly polarized beam.}\label{fig:Scheme}
\end{figure}

In this paper, we explore the mechanism of the generation of spin orientation in room-temperature Caesium vapour that combines three elements: (1) off-resonant optical pumping, (2) non-linear spin dynamics and (3) SEC (selective relaxation and coherence transfer \cite{Ruff1965, Haroche1970, Happer1977, Skalla1996, Chalupczak2014}). (1) A linearly polarized laser beam moves the atomic population from the F=3 to F=4 manifold through off-resonant optical pumping, while creating a population imbalance (alignment) within both levels, Fig.~\ref{fig:Scheme} (a).
The particular frequency detuning of the beam ensures that the majority of the population transferred to the F=4 level goes to either stretched states, i.e. sublevels with a maximum or minimum magnetic quantum number. While the dynamics within the F=3 level is defined by the resonant coupling to the laser field, the F=4 atomic spins evolve only in the presence of both weak far-off resonant optical and SEC couplings. (2) The weak coupling to the optical field drives the non-linear spin dynamics that breaks the population distribution symmetry. In particular, it moves some of the population out of one of the stretched states, effectively making the atoms more prone to SEC relaxation (3). As a consequence of these two factors (non-linear spin dynamics and SEC), we observe suppression of components representing one of the spin directions that contributes to alignment and the generation of atomic orientation at low magnetic fields. 
In contrast to standard alignment to orientation conversion experiments \cite{Cohen-Tannoudji1969, Lombardi1969, Hilborn1994, Budker2000, Kuntz2002, Auzinsh2010}, the change of the spin polarization discussed here is produced with a parallel magnetic and electric (linearly polarized laser beam) fields. In a sense, the mechanisms demonstrated here are a generalisation of that presented in \cite{Chalupczak2015}, where the major difference between the two experiments is the initial state created by the pumping process. While the indirect pumping implemented in \cite{Chalupczak2015} creates a system with 80\% of the population in the stretched state, here the starting point is an aligned state. 
This enables us to demonstrate a continuous transition from the aligned to oriented state with the scan of the Larmor frequency or the linearly polarized beam power.

The immediate implementation of the discussed technique is in the area of radio-frequency (rf) atomic magnetometry \cite{Savukov2005, Chalupczak2018}.
State-of-the-art rf atomic magnetometers take advantage of the properties of an oriented atomic polarization \cite{Savukov2005, Lee2006, Savukov2007, Chalupczak2012a}. This polarization ensures immunity of the magnetometer signal to SEC decoherence \cite{Appelt1999, Scholtes2011, Chalupczak2012} and enables a high signal-to-noise ratio (SNR) of the sensor.
Possible applications of ultrasensitive rf magnetic field sensors cover a wide range of technologies from chemical analysis \cite{Lee2006, Savukov2007, Bevilacqua2017} to the non-destructive testing of materials \cite{Wickenbrock2014, Bevington2018, Bevington2019, Bevington2019a, Bevington2019b}. A standard method for generating orientation in alkali metal atoms (optical pumping) uses a circularly polarized laser beam operating on the D1 line \cite{Ledbetter2008}. For even more efficient pumping, an additional laser, the so-called repumper, is added to transfer the atoms from the other hyperfine ground state \cite{Wasilewski2010}. In this configuration, the signal readout is done by a linearly polarized laser beam operating on the D2 line. Consequently, the magnetometer arrangement involves two or three lasers operating at different frequencies. The concept discussed here achieves a comparable sensor performance (SNR) with a single laser diode, which significantly simplifies the instrumentation.

The following part contains a brief description of the experimental instrumentation. The components of the atomic spin orientation mechanism are explored through the dependencies of the rf spectroscopy signal on three measurement parameters (laser frequency detuning, beam power and magnetic field strength), which are discussed in the subsequent sections.

\section{Experimental setup} 
The measurements are performed in a shielded environment \cite{Chalupczak2012a, Chalupczak2012, Chalupczak2014, Chalupczak2015}. The ambient magnetic field is suppressed by the use of five layers of cylindrical shields with end caps, made from 2-mm thick mu-metal. A solenoid inside the shield generates a well-controlled offset magnetic field, $B_{off}$, along $\hat{z}$ axis, Fig.~\ref{fig:Scheme} (b), with a relative homogeneity exceeding $10^{-4}$ over the length of the cell. The atoms used are a Caesium atomic vapour housed in a paraffin-coated, cross-shaped, quartz cell (22 mm in diameter and arm lengths of 32 mm)  at ambient temperature (atomic density $n_{\text{Cs}}=0.33 - 1.0\times10^{11} \text{cm}^{-3}$).
These atoms are optically pumped by a linearly polarised laser beam, 20 mm in diameter, propagating orthogonally to the direction of $B_{off}$,  Fig.~\ref{fig:Scheme} (b). The polarization of the beam is parallel to  $B_{off}$.  The beam is provided by a DBR diode laser operating on the caesium D2 line [Fig.~\ref{fig:Scheme} (a)] and can be frequency stabilized within $\pm \SI{10}{\giga\hertz}$ with respect to the master laser frequency using offset locking. The same linearly polarised beam also acts as a probe of the spin precession via the Faraday effect \cite{Takahashi1999}, where the evolution of the collective atomic spin is mapped onto the polarization state of the linearly polarized probe beam \cite{Savukov2005, Ledbetter2008, Wasilewski2010, Chalupczak2012a, Bevilacgua2016}. The laser light transmitted through the cell is analysed by a polarimeter consisting of a crystal polariser oriented at $45^{\circ}$ with respect to the incident polarization and a commercial balanced photodetector. The two quadrature components of the resulting signal X and Y (where X is the in-phase and Y is the out-off-phase components, R=$\sqrt{X^2+Y^2}$) are measured by a lock-in amplifier, referenced to the first harmonic of the driving rf field ($B_{rf}$) frequency.


\section{Off-resonant pumping} 
The change in population of the F=3 and F=4 ground states induced by the off-resonant pumping can be described by a pair of rate equations \cite{Atoneche2017}:
\begin{equation}
    \frac{d}{dt}G_{i} = G_j\sum_{F'}R_{j,F'}\beta_{j,F'}\beta_{i,F'} - G_{i}\sum_{F'}R_{i,F'}\beta_{i,F'}\left(1-\beta_{i,F'}\right),
    \end{equation}
where index $F'=2,3,4,5$ refers to the $6\,^2$P$_{3/2}$ excited state levels, the pair of indices $(i,j)$ equal $3$, and $4$ refer to the F=3 and F=4 $6\,^2$S$_{1/2}$ ground state levels and $\beta_{F,F'}$ is the coupling coefficient between the ground and excited state. The absorption rate $R_{F,F'}$ depends on the detuning of the pumping light from the relevant atomic transition. In our case, this dependence is dominated by Doppler broadening. Calculations based on the above equation show that the optimal condition (laser detuning) for population transfer from the F=3 to F=4 level and the generation of a population imbalance within the F=4 manifold are mutually exclusive. The former is optimised when the laser frequency is tuned in the vicinity of the $6\,^2$S$_{1/2}$ F=3$\rightarrow{}6\,^2$P$_{3/2}$ F'= 4 transition, at which the latter effect is minimised.

In the section below we experimentally identify the frequency range, which optimises the build-up of orientation in the F=4 level.

\begin{figure}[h!]
\includegraphics[width=\columnwidth]{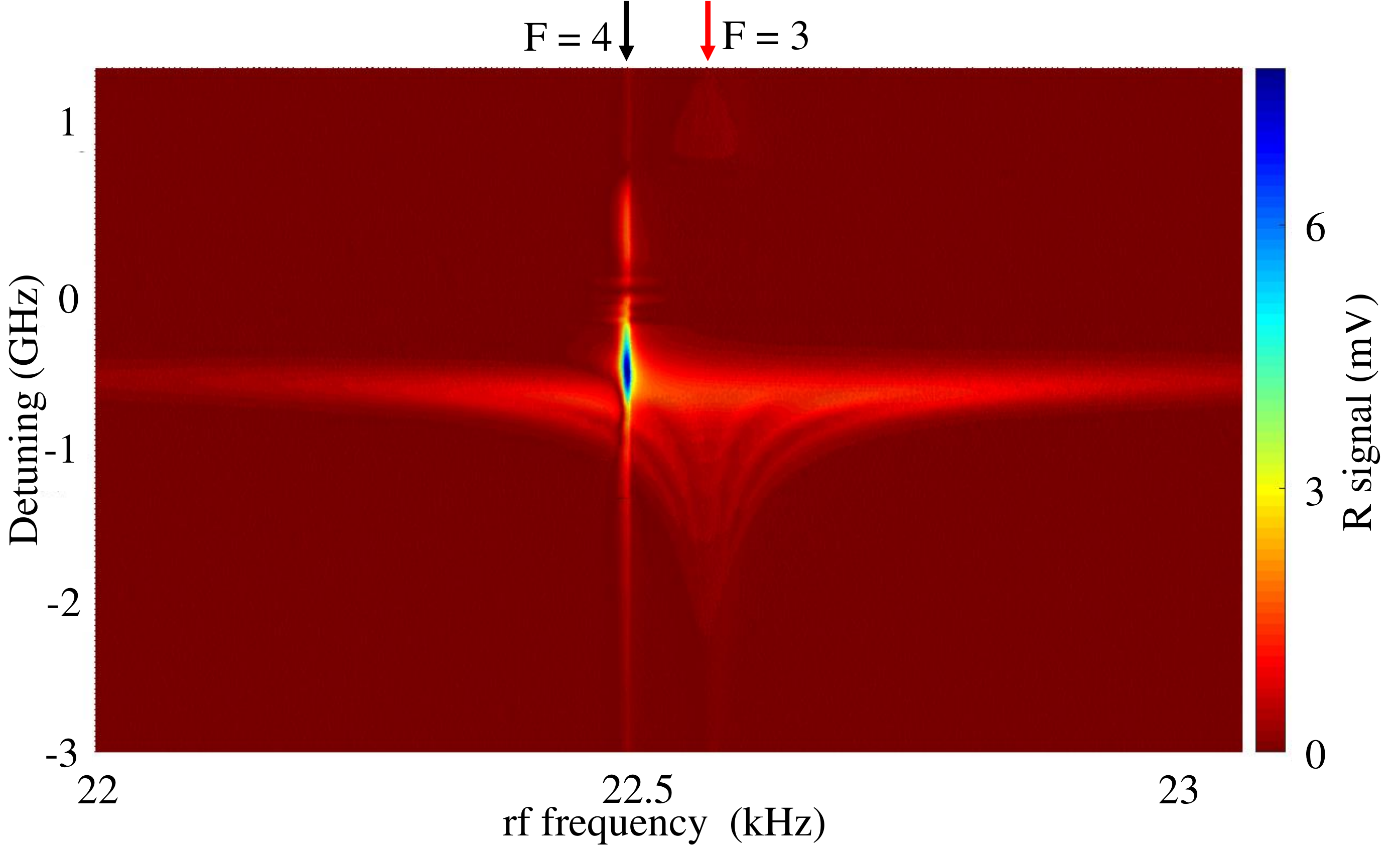}
\caption{ (colour online) Dependence of the rf signal amplitude, R, on the probe beam detuning from the $6\,^2$S$_{1/2}$ F=3$\rightarrow{}6\,^2$P$_{3/2}$ F'=2 transition. The F=3 and F=4 resonances are marked with a red and black arrow respectively. The measurements have been made with a laser beam power of 5.9 mW.}\label{fig:Detuning}
\end{figure}

Figure ~\ref{fig:Detuning} shows the rf signal magnitude, R, as the frequency of the linearly polarised laser beam is scanned across the group of D2 line transitions involving the F=3 ground-state (zero detuning represents the $6\,^2$S$_{1/2}$ F=3$\rightarrow{}6\,^2$P$_{3/2}$ F'=2 transition). The relatively small amplitude of $B_{off}$ (Larmor frequency $\sim \SI{22}{\kilo\hertz}$) ensures that, on one hand, the contributions from both ground-state levels can be individually distinguished and, on the other, the Zeeman levels in a particular manifold are degenerate. The splitting between components of the F=3 spectral profiles is defined solely by the tensor light shift. In particular, as the laser frequency approaches the atomic resonance, the tensor light shift in F=3 increases, and consequently, so does the splitting between the components of the relevant profile. Due to the relatively large detuning from resonance, there is no significant splitting in the F=4 profile. Efficient pumping from the F=3 to the F=4 level, in the vicinity of the $6\,^2$S$_{1/2}$ F=3$\rightarrow{}6\,^2$P$_{3/2}$ F'= 4 transition, results in the asymmetry in the F=3 signal amplitude with respect to the laser detuning. The exact detuning where the maximum in the F=4 signal is observed varies with the laser power and ranges from $\sim \SI{-416}{\mega\hertz}$ (3.3 mW) to $\sim \SI{-290}{\mega\hertz}$ (10 mW).
While the observation of off-resonant F=4 pumping is similar to that observed on the D1 transition \cite{Chalupczak2010}, there are two differences worth pointing out. Firstly, the maximum pumping between the manifolds is reached for a non-zero laser detuning ($\sim \SI{-310}{\mega\hertz}$ for the measurements represented in Fig. ~\ref{fig:Detuning}). Secondly, the character of the generated polarization within the F=3 (alignment) and F=4 (orientation) levels is different.

\begin{figure}[h!]
\includegraphics[width=\columnwidth]{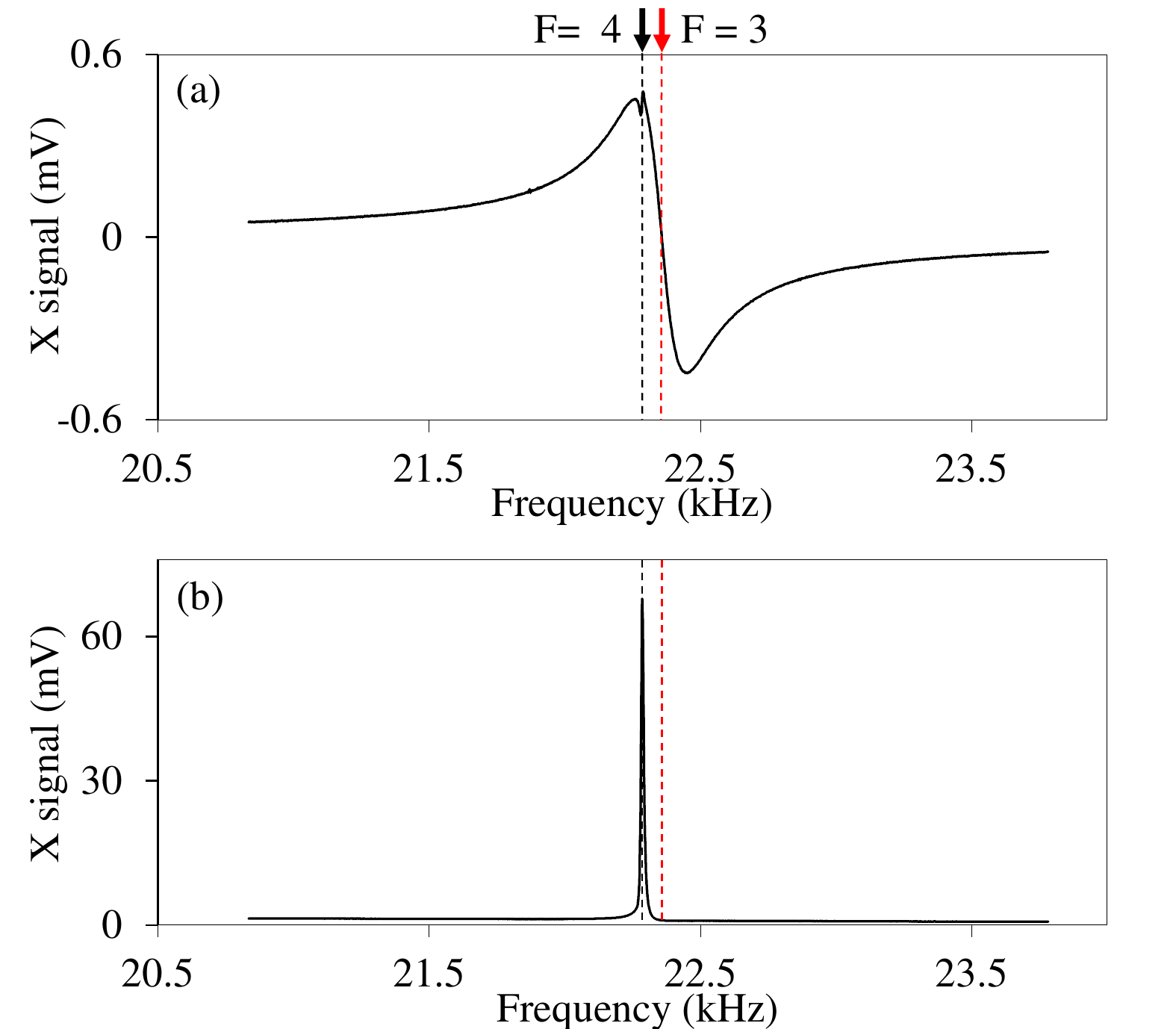}
\caption{The X component of the rf spectroscopy signal recorded with the laser beam frequency tuned near the $6\,^2$S$_{1/2}$ F=3$\rightarrow{}6\,^2$P$_{3/2}$ F'=2 transition (detuning -$\SI{100}{\mega\hertz}$). The F=3 and F=4 resonances are marked with a red and black arrow respectively. Transition from alignment (a) to orientation (b) can be seen in the spectral profile created by the F=4 coherences. The measurements were performed with a laser beam power of (a) 200 $\mu$W and (b) 9.1 mW.}\label{fig:F3spectra}
\end{figure}

\begin{figure}[ht!]
\includegraphics[width=\columnwidth]{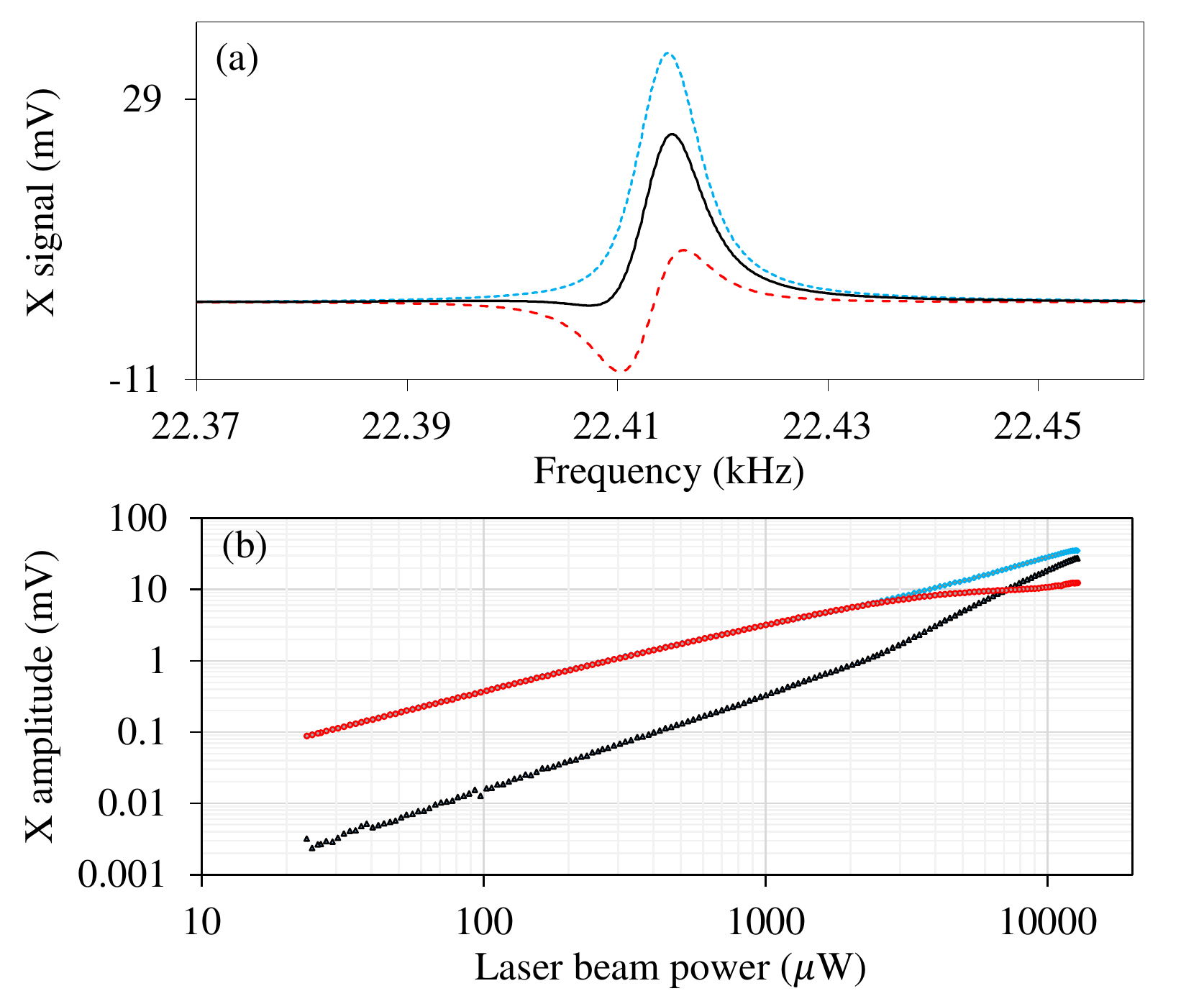}
\caption{(colour online) (a) Magneto-optical-rotation signal, represented by the X component of the rf spectroscopy signal, recorded with the linearly polarized beam only (solid black line), and in the presence of a circularly polarised pump beam with either orthogonal polarisation components (dashed red and pointed blue lines). The linearly polarized beam power is 12.4 mW and the pump power is $17 \mu$W. 
(b) Dependence of the X component's amplitude on the laser beam power for the linearly polarized beam only (black triangles) and combined with a circularly polarized beam parallel to $B_{off}$ (red points and blue diamonds represent the measurements with either of the two orthogonal circular polarisations for the pump beam).}\label{fig:AOC}
\end{figure}

\section{Non-linear dynamics}  
The F=4 atomic spins only evolve in the presence of weak far-off resonant optical and SEC couplings. The linearly polarized light ($\vec{E}_p$) couples to the atomic ground state through the tensor ac polarizability $\alpha_2$ (single-spin Hamiltonian, without the scalar part of the light shift $\sim \alpha_2(\vec{E}_p\cdot\hat{f}^{(i)})^2$, where $\hat{f}^{(i)}$ is the total angular momentum operator of the i'th atom). Therefore in general, the atomic spin dynamics will exhibit a non-linear character \cite{Smith2004, Chalupczak2015}. The topic of atomic spin evolution in the presence of an electric and magnetic field has been the subject of a number of theoretical \cite{Rochester2012} and experimental \cite{Budker2000, Kuntz2002, Smith2004, Chalupczak2015} studies. Mostly, the focus of these explorations was on measurement configurations where the two fields were orthogonal. In such a case, as shown in \cite{Rochester2012}, the atomic polarization exposed to a static magnetic and an off-resonant ac electric field begins to oscillate between two types of atomic polarization with an oscillation period equal to $\frac{2\pi}{\Delta\nu_{TLS}}=\frac{3}{2\hbar(2F-1)}\alpha_2|E_p|^2$, where $\Delta\nu_{TLS}$ is equal to the line separation generated by the tensor light shift. As pointed out in \cite{Smith2004} the non-linear dynamics critically depends on the relative angle between the fields, $\theta$, i.e. in the arrangement discussed here, the angle between the laser beam polarization and the $\hat{z}$ axis. In particular, the non-linear behaviour, quantified in \cite{Smith2004} by the spin decoherence, is twice as strong for an angle of $\theta=0^{\circ}$ than for $\theta=90^{\circ}$, while it disappears at $\theta=54^{\circ}$. Also, there has been a demonstration of an alkali-metal spin maser based on non-linear spin dynamics, which was performed at $\theta=0^{\circ}$ \cite{Chalupczak2015}. 
In the presence of SEC decoherence the behaviour of the system is defined by a combination of processes, which is characterised by the ratio of the $\Delta\nu_{TLS}$ and SEC relaxation rates.
Calculation of the spectral splitting \cite{Footnote} indicates that $\Delta\nu_{TLS}$ varies between $\SI{0.4}{\hertz}$ and $\SI{2}{\hertz}$ for a linearly polarised beam power of $2\,\text{mW}$ and 9$\,\text{mW}$ respectively. This is smaller than the relaxation rate due to SEC, $\SI{3}{\hertz}$ \cite{Chalupczak2014}. This indicates that the SEC process dominates the spin dynamics and while a small $\Delta\nu_{TLS}$ can trigger non-linear dynamics (similar to \cite{Chalupczak2015}), the system will not perform a full oscillation between atomic polarisation states. Intuitively, the mechanism is similar to the evolution of the atomic spin in a near-zero static magnetic field, where the combination of three effects (optical pumping, slow spin precession around the magnetic field and decoherence) effectively leads to a tilt of the steady-state spin polarization with respect to its initial state \cite{Kominis2003}. Here, the combination of (1) off-resonant pumping, which creates the alignment in the F=4 manifold, (2) non-linear spin dynamics that drives the alignment to orientation conversion, and (3) SEC decoherence results in a change (asymmetry) in the initial population distribution. The effect of conversion from alignment to orientation is enhanced by the SEC coherence transfer, which is discussed in the following section.

Figure ~\ref{fig:F3spectra} shows the X component of the rf spectra recorded with a laser power of 200 $\mu$W (a) and 9.1 mW (b). The positions of the F=3 and F=4 resonances are marked with red and black arrows respectively. Polarization-rotation resonances are observed when the rf field frequency matches the splitting between neighbouring Zeeman sublevels introduced by $B_{off}$.
In an aligned system, the rf response consists of two profiles with an opposite sign, leading to a dispersive-like lineshape. At low power, Fig. ~\ref{fig:F3spectra} (a), the rf spectrum consists of a large broad feature due to alignment in the F=3 level created by direct optical pumping, with a much smaller structure due to off-resonant excitation into the F=4 manifold.
An increase in the laser beam power does not only translate into an increase of the F=3 and F=4 signal amplitude, Fig. ~\ref{fig:F3spectra} (b). While the character of the F=3 profile remains unaltered, the change in the symmetry of the F=4 signal indicates the presence of atomic orientation.  The resonant coupling of laser light to the F=3 Zeeman sublevels results in power broadening of their corresponding spectral profiles, which contributes to the broad low amplitude background visible in Fig. ~\ref{fig:F3spectra} (b).

\begin{figure*}[ht]
\includegraphics[width=\textwidth]{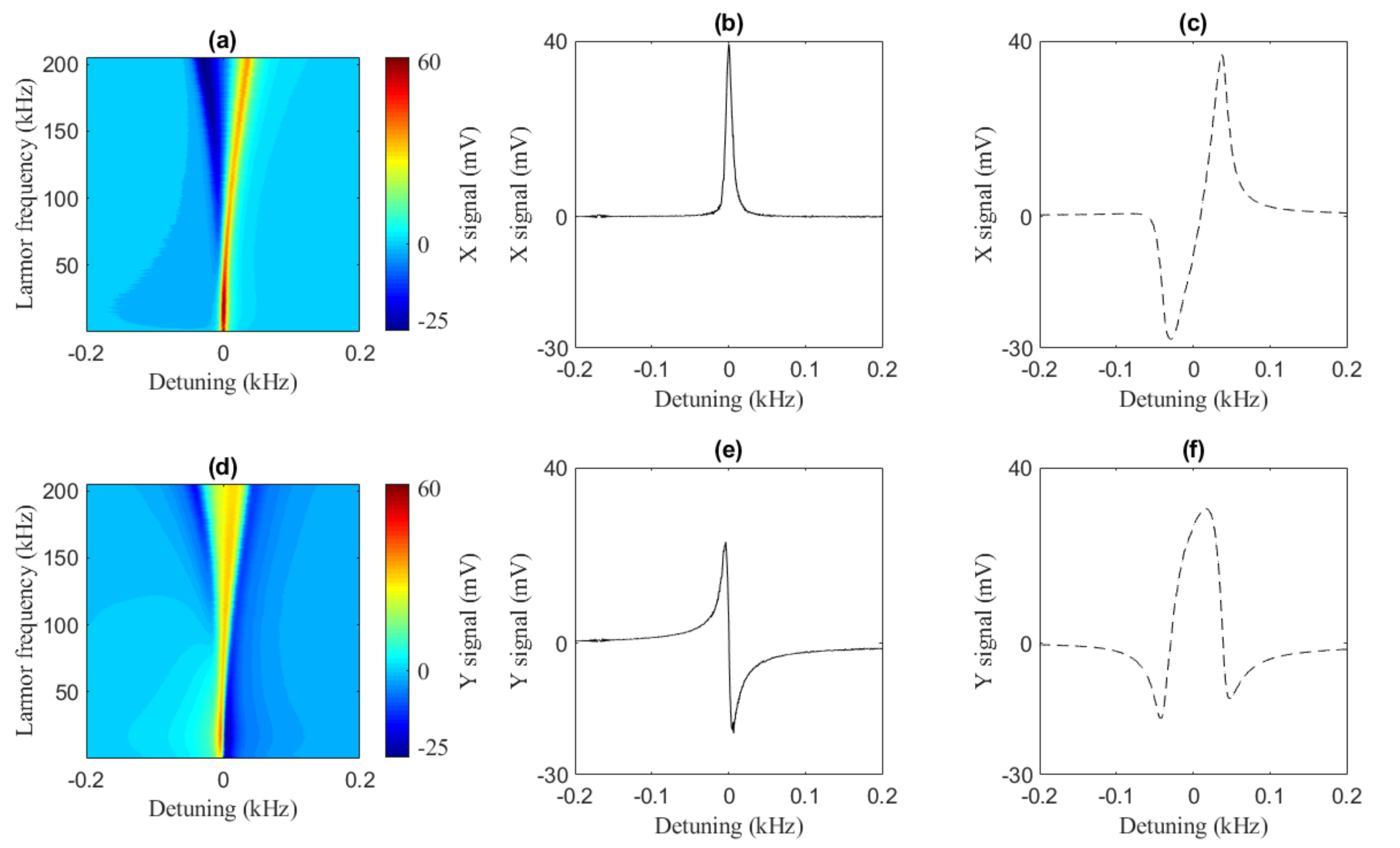}
\caption{(colour online) Dependence of the X (a) and Y (d) quadrature components of the rf signal on the offset magnetic field strength (Larmor frequency). The frequency of the rf spectra is expressed in terms of the detuning from the centre of the rf spectrum. Plots (a) and (e) are the X and Y components at a field strength equivalent to \SI{0.6}{\kilo\hertz}, while (c) and (f) are at \SI{204.7}{\kilo\hertz}. The measurements have been done with a laser beam power of 4.6 mW.}\label{fig:Magnetic_field}
\end{figure*}

To confirm that the F=4 spectral profile represents atomic orientation, pumping with a circularly polarized (pump) beam propagating along the direction of $B_{off}$ ($\hat{z}$ axis, Fig.~\ref{fig:Scheme}) was added \cite{Chalupczak2012}. The pump beam is generated by a diode laser, frequency locked to the Caesium $6\,^2$S$_{1/2}$ F=3$\rightarrow{}6\,^2$P$_{3/2}$ F'=2, 3 crossover. The solid black line in Fig. ~\ref{fig:AOC} (a) shows the rf spectrum for the F=4 profile recorded with only the linearly polarized beam. The dashed red and pointed blue lines represent the case where one of the two orthogonal circular polarisations of the pump beam is added. The presence of the pump beam creates atomic orientation in the sample (parallel or anti-parallel to $B_{off}$). 
If the orientation generated by the linearly polarised beam and pump beam coincide, the amplitude of the observed profile increases [pointed blue line in Fig. ~\ref{fig:AOC} (a)]. 
For the case with the opposite pump polarisation, the signal amplitude decreases and the character of the spectrum changes (dashed red line).
The dependencies of the signal amplitude for the opposite pump beam polarisations are shown (red points and blue diamonds ) in  Fig. ~\ref{fig:AOC} (b). The amplitudes of the signals created by the orthogonally polarized pump beams are equal below 2 mW of probe power. The asymmetry in amplitudes above this power is produced by the sample orientation induced by the linearly polarised beam. 
The signature of this effect is also present in the amplitude data for the signal produced by only the linearly polarised beam (black triangles). The change from a linear to a quadratic slope in the amplitude power dependence, seen above 2 mW, confirms the non-linear character of the underlying mechanism. 

\section{Spin-exchange collisions} 
The contribution of non-linear spin dynamics to the generation of orientation is enhanced by SEC driven coherence transfer \cite{Ruff1965, Haroche1970, Happer1977, Skalla1996, Chalupczak2014}. 
The effect relies on the macroscopic exchange of the excitation between different coherence modes subject to a resonance condition, which links the difference between the relevant coherence frequencies and the coherence relaxation rate \cite{Haroche1970}. If the size of the mismatch between the frequencies of the different modes is comparable to the SEC decoherence rate, rapid SECs average out the coherence precession with different Larmor frequencies leading to a prolonged oscillation at the averaged frequency \cite{Happer1977}. Intuitively, the resonant condition for coherence transfer is equivalent to an overlap of the spectroscopic profiles of the relevant coherences in the rf spectrum. 
For the case considered here, the degeneracy between the F=4 Zeeman sublevel transition frequencies leads to an automatic realisation of the resonance condition, i.e. the frequency mismatch (dephasing) between the precessing spins affected and not affected by SEC is negligible, so long as the SEC processes do not involve manifold change. Consequently, SEC processes do not contribute to relaxation and a reduction in the SEC dominated decoherence rate is observed \cite{Chalupczak2014}. 
One of the signatures of this coherence transfer is that the spectral profiles representing the relevant coherences group (merge) around the leading component of the spectrum \cite{Happer1977, Chalupczak2014}.

Figure ~\ref{fig:Magnetic_field} shows the dependence of the two normalised quadratures (X and Y) of the rf spectroscopy signal on $B_{off}$ (in terms of Larmor frequency). The normalization takes into account the variation of the amplitude and phase of the rf spectroscopy signal with operating frequency and was performed in the standard pump-probe configuration \cite{Chalupczak2012} over the same range of $B_{off}$. 
It is worth pointing out that SEC processes have a twofold contribution to the generation of orientation. Due to coherence transfer, tuning the strength of the offset magnetic field is equivalent to changing the value of the SEC relaxation rate. Reduction in the decoherence rate extends the duration of the non-linear alignment to orientation evolution. Collisions between atoms in their stretched state does not introduce relaxation as the total angular momentum must be conserved, therefore selective SEC relaxation depopulates all but the stretched state.
The spectral profile for large $B_{off}$ (Larmor frequency $ \geq \SI{100}{\kilo\hertz}$), Fig. ~\ref{fig:Magnetic_field} (c) and (f), have a shape typical for atomic alignment, Fig. ~\ref{fig:F3spectra} (a).
The decrease of $B_{off}$ increases the overlap of the components with opposite signs and consequently a reduction of the signal amplitude.
Non-linear spin dynamics results in a decrease in the population of the stretched state represented in Fig. ~\ref{fig:Magnetic_field} (a) and (c) by the profile positioned at $\sim \SI{-0.05}{\kilo\hertz}$ detuning from resonance. 
This results in a higher relaxation rate for the coherences contributing to the part of the spectrum with a negative detuning. Consequently, the amplitude of the component with a negative detuning decreases more rapidly with $B_{off}$ than the other profile. The small frequency mismatch between these coherence's (decreasing with the reduction of $B_{off}$) enhances coherence transfer and the build-up of atomic orientation observed over the frequency range below $\SI{20}{\kilo\hertz}$. The spectral profiles observed for Larmor frequencies $\leq \SI{20}{\kilo\hertz}$, Fig. ~\ref{fig:Magnetic_field} (b) and (e), represent the spin system in the oriented state. 

The maximum signal-to-noise ratio, recorded in the measurement conditions discussed here, is 1.3-1.4 times lower than that measured for the optimised indirect pumping scheme, \cite{Chalupczak2012}. Since the degree of population polarization described in \cite{Chalupczak2012}, i.e. stretched state population as a fraction of total population, was found to be 80\% we estimate the degree of polarization in the current configuration to be ~72\%.


\section{Conclusions} 
We have demonstrated the generation of atomic spin orientation in a room-temperature caesium vapour. The presence of atomic polarization is vital for the operation of a radio-frequency atomic magnetometer. 
The rf frequency range of the presented technique ($\SI{1}{\kilo\hertz}$-$\SI{30}{\kilo\hertz}$) is interesting in the context of magnetic induction based non-destructive testing, where a low operating frequency translates into a deeper penetration depth of the (so-called primary) magnetic field \cite{Bevington2018, Bevington2019, Bevington2019a, Bevington2019b}. The measurement configuration discussed here combines the efficient generation of the F=4 atomic orientation and off-resonant probing usually achieved with two/ three independent lasers. The obvious benefit of the presented scheme is the simplicity of the instrumentation. Systematic measurements of the signal-to-noise ratio (SNR) confirm that the discussed option delivers a SNR only 1.3-1.4 times lower than recorded in the optimised pump-probe configuration \cite{Chalupczak2012}. 
The relatively sharp peak in the signals frequency dependence, Fig. ~\ref{fig:Detuning}, allows for stabilisation of the laser frequency despite strong saturation of the F=3 resonance. 
The difficulty that the orientation generated by the linearly polarised beam is observed over a relatively narrow range of $B_{off}$ could be overcome through the implementation of a degenerate pump-probe configuration, which involves a circularly polarized (pump along the $\hat{z}$ axis) beam and a linearly polarized (probe along the $\hat{y}$ axis) beam operating at the same frequency ($\sim \SI{290}{\mega\hertz}$ from the $6\,^2$S$_{1/2}$ F=3$\rightarrow{}6\,^2$P$_{3/2}$ F'=2 transition). It is worth pointing out that this frequency for the pump beam is not far from that used in the optimised indirect pumping scheme, which means that a relatively small power, $\sim 100 \mu$W, will be required \cite{Chalupczak2012}. The Caesium ground state hyperfine splitting ($\SI{9.172}{\giga\hertz}$) defines the detuning of the laser frequency from F=4, which affects the signal amplitude and non-linearity strength. Hence, the use of $^{85}$Rb vapour (hyperfine splitting of $ \SI{3}{\giga\hertz}$) would improve sensor performance in two ways. (1) Smaller detuning of the light monitoring the atomic spin evolution will increase the recorded signal amplitudes and potentially improve the SNR. (2) The strength of the tensor couplings scales inversely with the square of the laser detuning \cite{Sherson2008}. Implementation of $^{85}$Rb vapour would reduce the linearly polarized beam power threshold for non-linear effects by nearly an order of magnitude ($\sim 200 \mu$W). 
Preliminary tests indicate that the combination of a degenerate pump-probe and the use of $^{85}$Rb vapour would enable efficient operation of the atomic magnetometer with 4 mW of laser light power, which is achievable from a single Vertical-Cavity Surface-Emitting Laser diode.

\section{Acknowledgements} 
\begin{acknowledgements}
The work was funded by UK Department for Business, Innovation and Skills as part of the National Measurement System Program.  P.B. was supported by the Engineering and Physical Sciences Research Council (EPSRC) (No. EP/P51066X/1).
\end{acknowledgements}

\end{document}